\documentclass{PoS}

\usepackage{amsmath,amssymb}
\usepackage{graphicx}
\usepackage{subfigure}

\title{Investigating a mixed action approach for $\eta$ and $\eta'$ mesons in $N_f=2+1+1$ lattice QCD}

\ShortTitle{$\eta$, $\eta'$ meson masses from a mixed action approach}

\author{Konstantin Ottnad, Carsten Urbach, \speaker{Falk Zimmermann}\\
  Institut f{\"u}r Strahlen- und Kernphysik, Rheinische Friedrich-Wilhelms-Universit{\"a}t Bonn\\
  Nussallee 14-16, 53115 Bonn \\ 
  \email{ottnad,urbach,fzimmermann@hiskp.uni-bonn.de}
}

\author{For the ETM Collaboration}

\abstract{
We present results for a test of a mixed action approach with
Osterwalder-Seiler valence quarks on a Wilson twisted mass sea for the
example $\eta$, $\eta'$ mesons. Flavour singlet pseudoscalar mesons
obtain significant contributions from disconnected diagrams and are,
therefore, expected to be particularly sensitive to mixed regularisations. We employ
different procedures for matching valence and sea quark actions and show that the results agree in the continuum limit.

}

\FullConference{31st International Symposium on Lattice Field Theory - LATTICE 2013\\
		July 29 - August 3, 2013\\
		Mainz, Germany}

\begin{document}

\section{Introduction}

{\it Mixed action} approaches, where valence and sea fermion actions are
chosen differently, are used frequently in lattice QCD. They possess a
number of important advantages compared to the so called {\it unitary}
case, where valence and sea quark actions are identical. In
particular, it is possible to use a valence action
obeying more symmetries than the sea action in cases where the valence
action cannot be used in the sea for theoretical reasons or because of
too high computational costs. One can even go one step further and try
to correct for small mismatches in bare parameters in the sea
simulation by using a partially quenched mixed action
approach. 


Of course, a mixed action approach has also disadvantages, most
prominently the breaking of unitarity, which might for instance drive
certain correlators negative. Also, it is not clear a priori how big
lattice artifacts one encounters in mixed formulations. 

In this proceeding contribution we will present results on a mixed action approach with so called Osterwalder-Seiler~\cite{Frezzotti:2004wz} valence quarks on a $N_f=2+1+1$ flavour Wilson twisted mass sea~\cite{Frezzotti:2000nk} and compare to unitary results~\cite{Ottnad:2012fv,Michael:2013gka,konstantin:proceeding13}.
This particular action combination has the advantage that the flavour symmetry breaking present in the sea formulation is avoided in the valence formulation. Moreover, the matching of valence and sea actions is particularly simple~\cite{Frezzotti:2004wz}.
As physical example we study the $\eta$ and $\eta'$ system.
The corresponding correlation functions obtain significant contributions from disconnected diagrams and are, therefore, uniquely sensitive to differences in between valence and sea formulations\footnote{Note that this was also discussed in the context of the validity of the fourth root trick in staggered simulations, see Refs.~\cite{Gregory:2006wk,Gregory:2011sg} and references therein.}.
We study the continuum limit with different matching conditions and find remarkably good agreement to the unitary case. First accounts of this work can be found in Ref.~\cite{Cichy:2012hq}.



\section{Lattice actions}
\label{sec:actions}

The results we will present in this proceeding are obtained by
evaluating gauge configuration provided by the European Twisted Mass
Collaboration (ETMC)~\cite{Baron:2010bv,Baron:2010th,Baron:2011sf}. We use the ensembles
specified in table~\ref{tab:setup} adopting the notation from
Ref.~\cite{Baron:2010bv}. More details can be found in this
reference. All errors are computed using a blocked bootstrap procedure
to account for autocorrelation.

The sea quark formulation is the Wilson twisted mass formulation with
$N_f=2+1+1$ dynamical quark flavours. The Dirac operator for $u$ and
$d$ quarks reads 
~\cite{Frezzotti:2000nk} 
\begin{equation}
 D_\ell = D_W + m_0 + i \mu_\ell \gamma_5\tau^3\, ,
 \label{eq:Dlight}
\end{equation}
where $D_W$  denotes the standard Wilson operator and $\mu_\ell$ the
bare light twisted mass parameter. For the heavy doublet of $c$ and
$s$ quarks~\cite{Frezzotti:2003xj} the Dirac operator is given by
\begin{equation}
 D_h = D_W + m_0 + i \mu_\sigma \gamma_5\tau^1 + \mu_\delta \tau^3\,,
 \label{eq:Dsc}
\end{equation}
with $\tau^i$ representing the Pauli matrices in flavour space.
We remark that the twisted term introduces flavour mixing between
strange and charm quarks that needs to be taken into account in the
analysis. 

The bare Wilson quark mass $m_0$ has been tuned to its critical
value $m_\mathrm{crit}$~\cite{Chiarappa:2006ae,Baron:2010bv} to achieve
automatic order $\mathcal{O}\left(a\right)$ improvement~\cite{Frezzotti:2003ni}, which is one of the main advantages of this formulation. 

In the valence sector we employ the Osterwalder-Seiler (OS)
action~\cite{Frezzotti:2004wz}. Formally, we introduce a twisted
doublet both for valence strange and charm
quarks~\cite{Frezzotti:2004wz,Blossier:2007vv}. The Dirac operator
for a single quark flavour $q\in \{ s,c \}$
\begin{equation}
 D_{q,q'} = D_W + m_\mathrm{crit} \pm i \mu_q \gamma_5
\end{equation}
is formally identical to the one in the light sector Eq. \ref{eq:Dlight}. Flavour $q$ ($q'$) will come with $+\mu_q$
($-\mu_q$). We denote the strange (charm) quark mass with
$\mu_s$ ($\mu_c$). In Ref.~\cite{Frezzotti:2004wz} it was shown that
automatic $\mathcal{O}(a)$-improvement stays valid and unitarity is
restored in the continuum limit.

For matching the strange quark mass we employ two procedures based on
meson masses. In previous studies it was found that matching kaon
masses is best in the sense that the residual lattice artifacts in the
results computed in a mixed framework are small~\cite{Sharpe:2004ny}. The
corresponding interpolating operator in the OS framework reads
\begin{equation}
  \label{eq:OK+}
  \mathcal{O}_{K^{+}} =\bar\psi_s\ i\gamma_5\ \psi_d\,.
\end{equation}
Note that we rely to the so called physical basis throughout this
proceeding contribution~\cite{Shindler:2007vp}. For details on how to compute the
kaon mass in the unitary case we refer to Ref.~\cite{Baron:2010th}.
As a second matching observable we use the mass $M_{\eta_s}$ of the so-called $\eta_s$ meson -- a pion
made out of strange quarks which does not exist in nature. $M_{\eta_s}$ can be obtained from the connected only correlation function of the OS interpolating operator
\begin{equation}
  \label{eq:etas}
  \mathcal{O}_{\eta_s} =\frac{1}{\sqrt{2}}(\bar\psi_s\ i\gamma_5\
  \psi_s + \bar\psi_{s'}\ i\gamma_5\ \psi_{s'})\,. 
\end{equation}
In both cases we tune the value of $a\mu_s$ such that the kaon
($\eta_s$) masses agree within error in between the mixed and the
unitary formulation. In order to compute the matching values
for $\mu_s$ we performed inversions in a range of $a\mu_s$ values
around a first guess obtained from $\mu_s =
\mu_\sigma-Z_P/Z_S\mu_\delta$, computed $M_{K^+}$ and $M_{\eta_s}$, and
interpolated to the matching point where needed. The matching values
for $a\mu_s$ for the two matching observables and all ensembles can be
found in table~\ref{tab:setup}.

The value of the charm quark mass turns out to be not very important
for our investigation, because the charm does not contribute
significantly to $\eta$ and $\eta'$ mesons. Therefore, we use the
following relations for the bare twisted quark mass parameters to
obtain 
\begin{equation}
 \label{eq:musc}
 \mu_{c} = \mu_\sigma\ +\ Z_P/Z_S\ \mu_\delta \,.
\end{equation}
The value for the ratio of renormalisation constants 
can be found in Ref.~\cite{ETM:2011aa}. The actual values for $\mu_c$ can again
be found in table~\ref{tab:setup}.

\begin{table}[t!]
 \centering
 \begin{tabular*}{1.\textwidth}{@{\extracolsep{\fill}}lcccc|cccccc}
  \hline\hline
  ensemble & $\beta$ & $a\mu_\ell$ & $a\mu_\sigma$ & $a\mu_\delta$ & $a\mu^{K^+}_s$ & $a\mu^{\eta_s}_s$ & $a\mu_c$ & $L/a$ & $N_\mathrm{conf}$ \\ 
  \hline\hline
  $A60.24$   & $1.90$ & $0.0060$ & $0.150$  & $0.190$  & $0.0232$ & $0.0138$ & $0.2768$ & $24$ & $1177$    \\
  $B55.32$   & $1.95$ & $0.0055$ & $0.135$  & $0.170$  & $0.0186$ & $0.0110$ & $0.2514$ & $32$ & $1964$   \\
  $D45.32sc$ & $2.10$ & $0.0045$ & $0.0937$ & $0.1077$ & $0.0149$ & $0.0118$ & $0.1720$ & $32$ & $885$   \\
  \hline\hline
  \vspace*{0.1cm}
 \end{tabular*}
 \caption{The ensembles used in this investigation. For the labeling
   we employ the notation of ref.~\cite{Baron:2010bv}. Additionally,
   we give the number of the configurations $N_\mathrm{conf}$ which were used to compute correlators in light and strange quark sector, the mass of the OS valence strange quark $\mu_s$ for $M_{K^+}$ and $M_{\eta_s}$ matching and the OS valence charm quark mass $\mu_c$. }
 \label{tab:setup}
\end{table}

\section{Pseudoscalar flavour-singlet mesons}

In order to extract $\eta$ and $\eta'$ states we 
compute the Euclidean correlation functions 
\begin{equation}
  \label{eq:correlations}
  \mathcal{C}(t)_{qq'} =
  \langle\mathcal{O}_q(t'+t)\mathcal{O}_{q'}(t')\rangle\,,\quad
  q,q'\in\{u,s,c\}\,, 
\end{equation}
with operators $\mathcal{O}_q = (\bar qi\gamma_5 q + \bar q'
i\gamma_5 q')/\sqrt{2}$, with $q\in \{u,s,c\}$ and identifying $u'\equiv
d$, again relying to the physical basis. The generalised eigenvalue
problem is solved~\cite{Michael:1982gb,Luscher:1990ck,Blossier:2009kd} for
determining $M_\eta$ and $M_{\eta'}$.
Correlation functions are made of quark connected diagrams and disconnected quark loops.
The quark connected pieces have been calculated via the so called  ``one-end-trick'' \cite{Boucaud:2008xu} using 
stochastic timeslice sources and for the disconnected diagrams we resort
to stochastic volume sources with complex Gaussian noise.

In the twisted mass formulation a very powerful variance
reduction method is available for estimating the disconnected loop
$(\bar ui\gamma_5 u + \bar d'i\gamma_5 d')/\sqrt{2}$, see
Ref.~\cite{Jansen:2008wv}. In the OS case this variance reduction also applies to
strange and charm disconnected loops~\cite{Cichy:2012hq}. Double
counting of loops stemming from this approach needs to be taken into
account by combinatorial factors. 

Finally, we can define mixing angles $\phi_l$, $\phi_s$ in the quark
flavour basis using the pseudoscalar matrix elements, see Ref.~\cite{Ottnad:2012fv}.
From chiral perturbation theory combined with large $N_C$ arguments
$|\phi_l-\phi_s|/|\phi_l+\phi_s|\ll1$ can be inferred according to
Refs.~\cite{Kaiser:1998ds,Kaiser:2000gs,Feldmann:1998sh,Feldmann:1998vh}
which is confirmed by lattice QCD~\cite{Michael:2013gka}.
Therefore, we will consider only the average mixing angle
  \begin{equation}
    \tan^2\phi \equiv
        \frac{A_{l,\eta'}A_{s,\eta}}{A_{l,\eta}A_{s,\eta'}} \,,
  \end{equation}
  where the $A_{q,n} = \langle 0 | \bar q i\gamma_5 q|n\rangle$ are pseudoscalar
  matrix elements determined from the eigenvectors with $n \in  \{\eta ,\ \eta'\}$ and $q \in \{l,\ s\}$.

\subsection*{Excited State Removal}

To improve the $\eta'$ (and $\eta$) mass determinations, we use a
method first proposed in Ref.~\cite{Neff:2001zr}, successfully
applied for the $\eta_2$ (the $\eta'$ in $N_f=2$ flavour QCD) in
Ref.~\cite{Jansen:2008wv} and very recently to the $N_f=2+1+1$ case in
Ref.~\cite{Michael:2013gka}. It grounds on the assumption that
disconnected contributions are significant only for the $\eta$ and
$\eta'$ state, but negligible for higher excited states. The method
involves to subtract excited states from the connected correlators
only. The subtracted connected and full disconnected are combined in 
$\mathcal{C}$, which is then used in the analysis. We refer to the
discussion in Ref.~\cite{konstantin:proceeding13,Michael:2013gka} for more details. 
%

\section{Results}

In order to compare the mixed case with the unitary case we match the
two actions as detailed in the previous sections using either the kaon
or the $\eta_s$ mass. Next we compute $M_\eta^\mathrm{OS}$ at this
matching points and the difference to the corresponding unitary masses
$\Delta M_\eta$. At fixed values of $r_0 M_\pi$ and $r_0 M_K$ -- which
is approximately the case for the ensembles D45, B55 and A60,
$r_0\Delta M_\eta$ should go to zero in the continuum limit with a
rate of $\mathcal{O}(a^2)$. We do not expect the small differences in
$r_0 M_K$ in between the different lattice spacing values to influence
our results much, because unitary and OS data are affected likewise.

\begin{figure}[t]
  \centering
  \subfigure[]{\includegraphics[height=6cm]{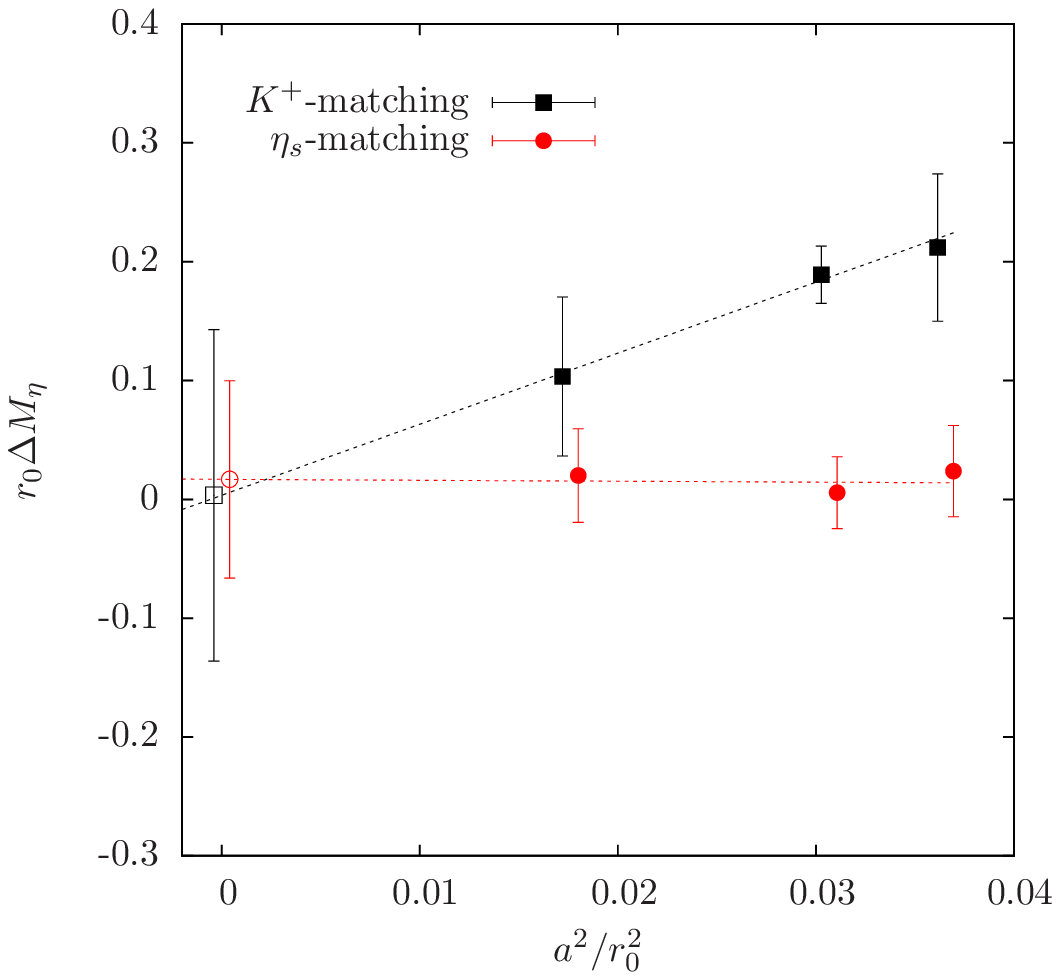}}\quad
  \subfigure[]{\includegraphics[height=6cm]{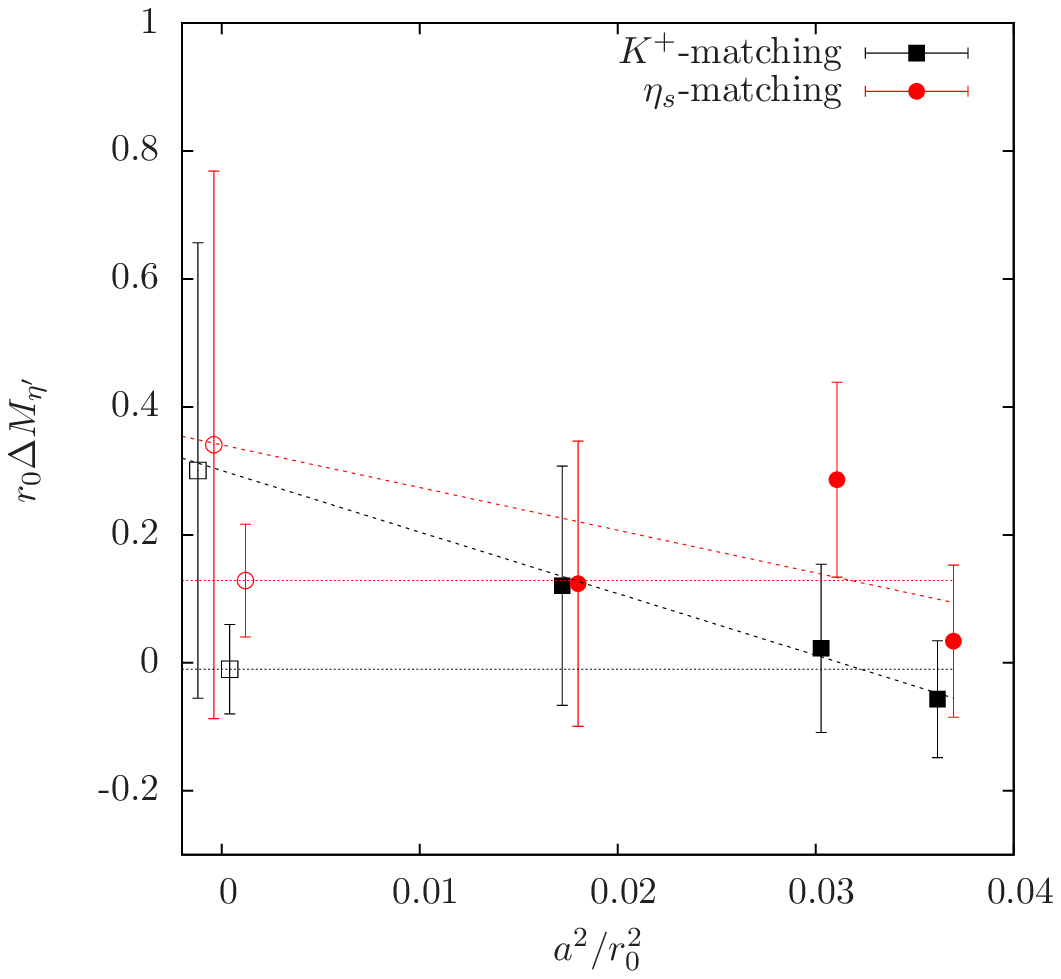}}
   \caption{(a) We show  $\Delta M_\eta$ for three ensembles (D45.32sc, B55.32, A60.24)
  as a function of the squared lattice spacing. Leftmost points correspond to linear continuum extrapolations indicated by the lines.
     (b) the same as (a) but for the $\eta'$ state, for which we additionally show constant extrapolations.
   }
  \label{fig:compOS}
\end{figure}

\begin{figure}[t]
  \centering
  \subfigure[]{\includegraphics[height=6cm]{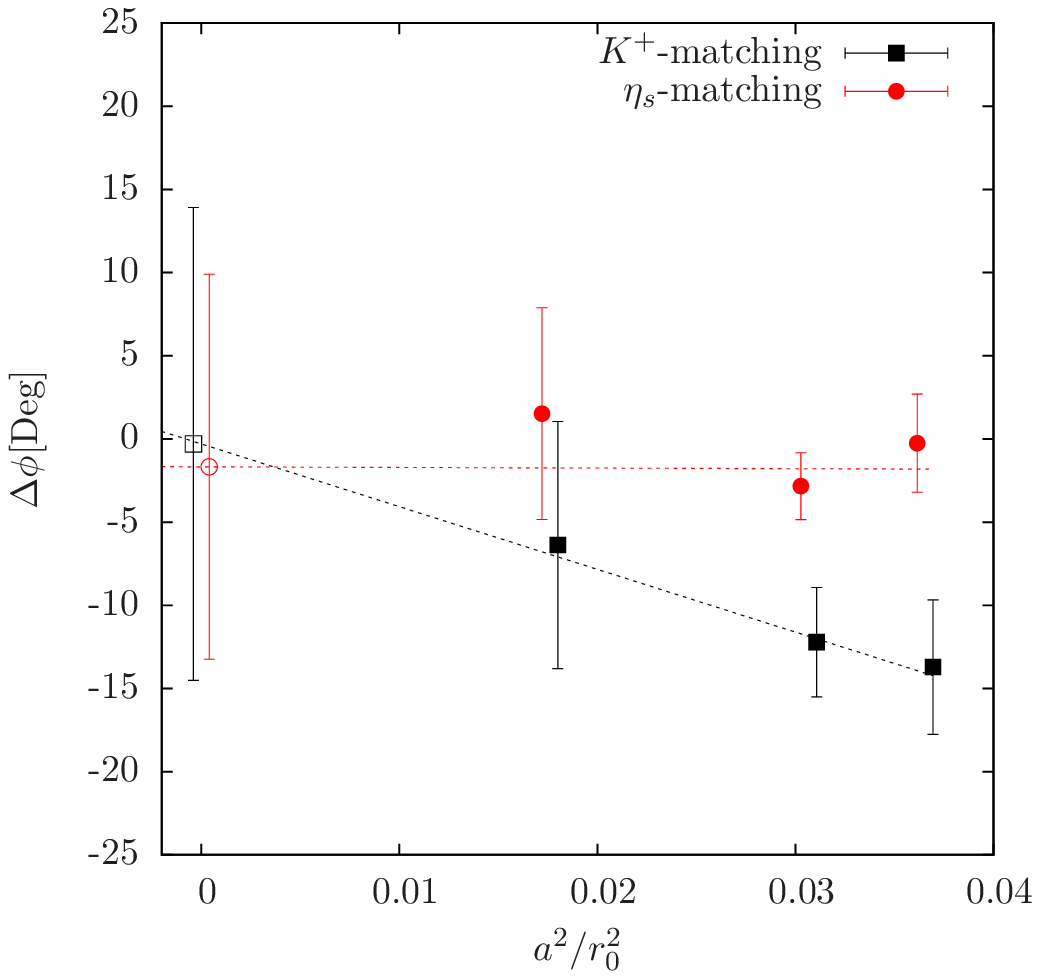}}
  \subfigure[]{\includegraphics[height=6cm]{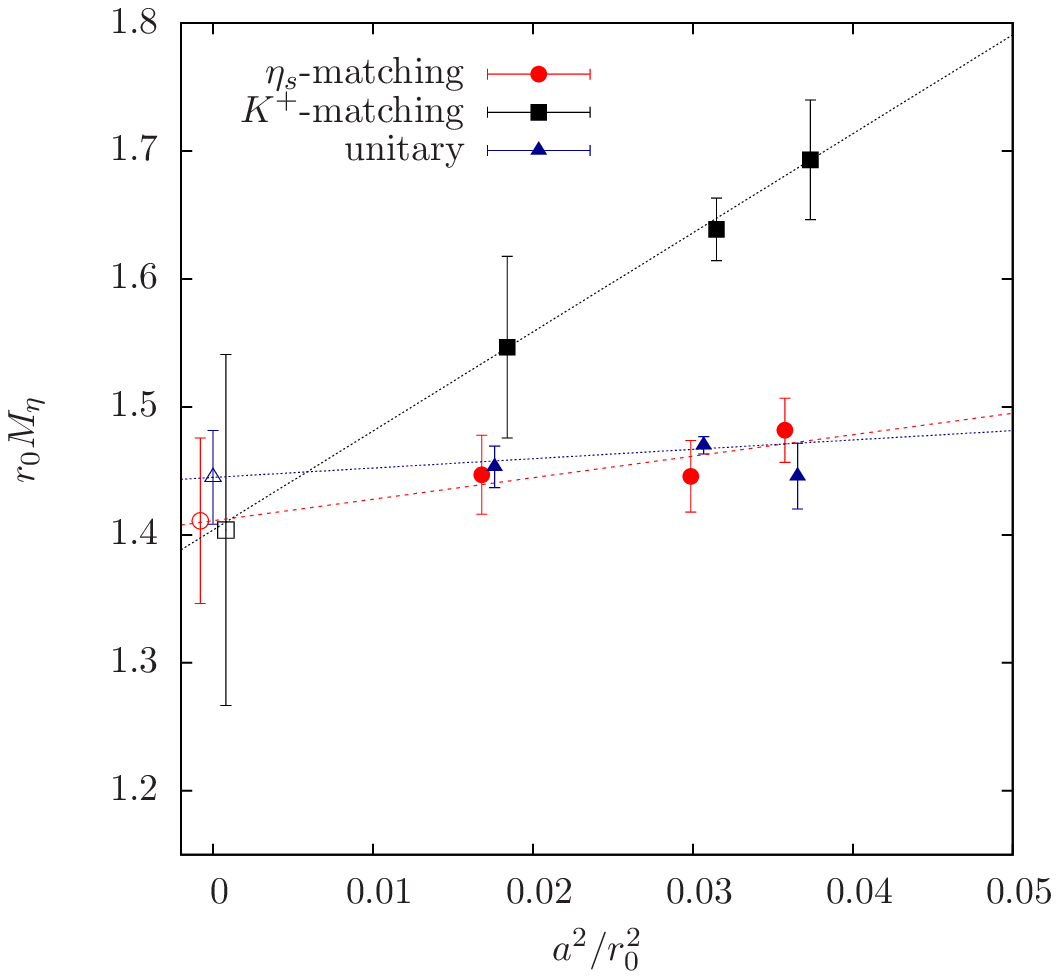}}\quad
  \caption{
    (a) We show $\Delta\phi$ for three ensembles (D45.32sc, B55.32, A60.24) as a function of the squared lattice spacing. Leftmost points correspond to linear continuum extrapolations indicated by the lines.
    (b) $r_0 M_\eta$ as a function of $(a/r_0)^2$ for the three ensembles.
      Different symbols correspond to two OS results
     with different matching conditions and to the unitary results. The
     leftmost points represent the corresponding continuum extrapolated
     values.
  }
  \label{fig:matched}
\end{figure}

$r_0\Delta M_\eta$ is shown in the left panel of
figure~\ref{fig:compOS} as a function of $a^2/r_0^2$. For both
matching observables we observe a linear dependence in $a^2/r_0^2$. A
corresponding continuum extrapolation in $a^2/r_0^2$ leads to the
expected vanishing of this difference at $a=0$. Kaon matching clearly
exhibits larger differences, while $\eta_s$ matching gives
$r_0\Delta M_\eta$ compatible with zero for each value of the lattice
spacing separately. This indicates smaller $a^2$ artifacts for $\eta_s$ matching because the unitary masses show constant scaling in $a^2$~\cite{Ottnad:2012fv}.
 $r_0 M_{\eta'}$ is shown in figure~\ref{fig:compOS}(b), which is in most cases compatible with zero even at finite lattice spacing.
 In the left panel of figure~\ref{fig:matched} we 
show $\Delta\phi$, again for both matching procedures, with the same
conclusion. 

As discussed in the introduction, one can correct for small mismatches
in the bare simulation parameters used for the sea action by going
slightly partially quenched in the valence sector. In order to test
also this approach, we have tuned the respective $a\mu_s$ values for
the ensembles D45, B55 and A60 such that all ensembles yield the same values of
$r_0M_K = 1.341$ or $r_0M_{\eta_s}=1.571$ .

At these matching points we again compute $M_\eta$. In the right panel of figure~\ref{fig:matched} we show $r_0M_\eta$ as a function of $a^2/r_0^2$ for
both matching conditions.
We observe that, within statistical uncertainties, both matching procedures lead to the same continuum results. 


A direct comparison to the unitary case is possible by also
correcting the unitary values of $M_\eta$ for the small mismatch in
the bare strange quark mass values. This is possible by measuring
$dM_\eta^2/dM_K^2$ and use it to shift all $M_\eta$ values to the same
value of $r_0 M_K=1.341$~\cite{Ottnad:2012fv}. The result is again shown in the right panel of
figure~\ref{fig:matched}. The corresponding continuum extrapolation is linear in $a^2/r_0^2$ and also compatible to both OS continuum points within errors.

\section{Conclusions and Outlook}

We tested a mixed action approach with Osterwalder-Seiler valence
quarks on a Wilson twisted mass sea for the $\eta$ and $\eta'$
system.
We expect this system to be most sensitive to different valence and sea discretisation due to significant disconnected contributions in the correlation functions. 

We employed two different conditions to match valence and sea actions,
using the kaon mass and the $\eta_s$ meson mass. For both matching
procedures we observe agreement in the continuum limit, and in
particular also with the unitary results. Moreover, we have also
corrected for small mismatches in the bare strange quark mass value
used in the gauge configuration generation, leading to a partially
quenched mixed action approach. Also in this case we found within errors identical
continuum extrapolated values for $M_\eta$ and the mixing angle $\phi$
for both matching procedures. 

By correcting also the unitary values of $M_\eta$ we found agreement to the OS results. Therefore, we conclude that the
mixed action approach can be used in the delicate case of the $\eta$
and $\eta'$ meson, at least to the precision that we could achieve
here. 

We thank all members of ETMC for the most enjoyable collaboration. The
computer time for this project was made available to us by the John von
Neumann-Institute for Computing (NIC) on the JUDGE and Jugene
systems. In particular we thank U.-G.~Mei{\ss}ner for granting us access on JUDGE. This project was funded by the
DFG as a project in the SFB/TR 16. K.O. and C.U. were supported by
the BCGS of Physics and Astronomie. The open source software packages tmLQCD~\cite{Jansen:2009xp} and R~\cite{R:2005} have been used. 

\bibliographystyle{h-physrev5}
\bibliography{bibliography}

\begin{thebibliography}{10}

\bibitem{Frezzotti:2004wz}
R.~Frezzotti and G.~C. Rossi,
\newblock JHEP {\bf 10}, 070 (2004),
  \href{http://arxiv.org/abs/hep-lat/0407002}{{\tt arXiv:hep-lat/0407002}}.

\bibitem{Frezzotti:2000nk}
{\bf ALPHA} Collaboration, R.~Frezzotti, P.~A. Grassi, S.~Sint and P.~Weisz,
\newblock JHEP {\bf 08}, 058 (2001),
  \href{http://arxiv.org/abs/hep-lat/0101001}{{\tt hep-lat/0101001}}.

\bibitem{Ottnad:2012fv}
{\bf ETM} Collaboration, K.~Ottnad {\em et~al.},
\newblock JHEP {\bf 1211}, 048 (2012),
  \href{http://arxiv.org/abs/1206.6719}{{\tt arXiv:1206.6719}}.

\bibitem{Michael:2013gka}
C.~Michael, K.~Ottnad and C.~Urbach,
\newblock Phys. Rev. Lett. 111, {\bf 181602} (2013),
  \href{http://arxiv.org/abs/1310.1207}{{\tt arXiv:1310.1207}}.

\bibitem{konstantin:proceeding13}
K.~Ottnad {\em et~al.},
\newblock PoS {\bf LATTICE2013}, 253 (2013).

\bibitem{Gregory:2006wk}
E.~B. Gregory, A.~C. Irving, C.~M. Richards and C.~McNeile,
\newblock PoS {\bf LAT2006}, 176 (2006),
  \href{http://arxiv.org/abs/hep-lat/0610044}{{\tt arXiv:hep-lat/0610044}}.

\bibitem{Gregory:2011sg}
{\bf UKQCD} Collaboration, E.~B. Gregory, A.~C. Irving, C.~M. Richards and
  C.~McNeile,
\newblock Phys.Rev. {\bf D86}, 014504 (2012),
  \href{http://arxiv.org/abs/1112.4384}{{\tt arXiv:1112.4384}}.

\bibitem{Cichy:2012hq}
K.~Cichy {\em et~al.},
\newblock PoS {\bf LATTICE2012}, 151 (2012),
  \href{http://arxiv.org/abs/1211.4497}{{\tt arXiv:1211.4497}}.

\bibitem{Baron:2010bv}
{\bf ETM} Collaboration, R.~Baron {\em et~al.},
\newblock JHEP {\bf 06}, 111 (2010), \href{http://arxiv.org/abs/1004.5284}{{\tt
  arXiv:1004.5284}}.

\bibitem{Baron:2010th}
{\bf ETM} Collaboration, R.~Baron {\em et~al.},
\newblock Comput.Phys.Commun. {\bf 182}, 299 (2011),
  \href{http://arxiv.org/abs/1005.2042}{{\tt arXiv:1005.2042}}.

\bibitem{Baron:2011sf}
R.~Baron {\em et~al.},
\newblock PoS {\bf LATTICE2010}, 123 (2010),
  \href{http://arxiv.org/abs/1101.0518}{{\tt arXiv:1101.0518}}.

\bibitem{Frezzotti:2003xj}
R.~Frezzotti and G.~C. Rossi,
\newblock Nucl. Phys. Proc. Suppl. {\bf 128}, 193 (2004),
  \href{http://arxiv.org/abs/hep-lat/0311008}{{\tt hep-lat/0311008}}.

\bibitem{Chiarappa:2006ae}
T.~Chiarappa {\em et~al.},
\newblock Eur. Phys. J. {\bf C50}, 373 (2007),
  \href{http://arxiv.org/abs/hep-lat/0606011}{{\tt arXiv:hep-lat/0606011}}.

\bibitem{Frezzotti:2003ni}
R.~Frezzotti and G.~C. Rossi,
\newblock JHEP {\bf 08}, 007 (2004),
  \href{http://arxiv.org/abs/hep-lat/0306014}{{\tt hep-lat/0306014}}.

\bibitem{Blossier:2007vv}
{\bf ETM} Collaboration, B.~Blossier {\em et~al.},
\newblock JHEP {\bf 04}, 020 (2008), \href{http://arxiv.org/abs/0709.4574}{{\tt
  arXiv:0709.4574}}.

\bibitem{Sharpe:2004ny}
S.~R. Sharpe and J.~M. Wu,
\newblock Phys.Rev. {\bf D71}, 074501 (2005),
  \href{http://arxiv.org/abs/hep-lat/0411021}{{\tt arXiv:hep-lat/0411021}}.

\bibitem{Shindler:2007vp}
A.~Shindler,
\newblock Phys.Rept. {\bf 461}, 37 (2008),
  \href{http://arxiv.org/abs/0707.4093}{{\tt arXiv:0707.4093}}.

\bibitem{ETM:2011aa}
{\bf ETM} Collaboration, B.~Blossier {\em et~al.},
\newblock PoS {\bf LATTICE2011}, 233 (2011),
  \href{http://arxiv.org/abs/1112.1540}{{\tt arXiv:1112.1540}}.

\bibitem{Michael:1982gb}
C.~Michael and I.~Teasdale,
\newblock Nucl.Phys. {\bf B215}, 433 (1983).

\bibitem{Luscher:1990ck}
M.~L{\"uscher} and U.~Wolff,
\newblock Nucl.Phys. {\bf B339}, 222 (1990).

\bibitem{Blossier:2009kd}
B.~Blossier, M.~Della~Morte, G.~von Hippel, T.~Mendes and R.~Sommer,
\newblock JHEP {\bf 0904}, 094 (2009),
  \href{http://arxiv.org/abs/0902.1265}{{\tt arXiv:0902.1265}}.

\bibitem{Boucaud:2008xu}
{\bf ETM} Collaboration, P.~Boucaud {\em et~al.},
\newblock Comput.Phys.Commun. {\bf 179}, 695 (2008),
  \href{http://arxiv.org/abs/0803.0224}{{\tt arXiv:0803.0224}}.

\bibitem{Jansen:2008wv}
{\bf ETM} Collaboration, K.~Jansen, C.~Michael and C.~Urbach,
\newblock Eur.Phys.J. {\bf C58}, 261 (2008),
  \href{http://arxiv.org/abs/0804.3871}{{\tt arXiv:0804.3871}}.

\bibitem{Kaiser:1998ds}
R.~Kaiser and H.~Leutwyler,
\newblock \href{http://arxiv.org/abs/hep-ph/9806336}{{\tt
  arXiv:hep-ph/9806336}}.

\bibitem{Kaiser:2000gs}
R.~Kaiser and H.~Leutwyler,
\newblock Eur.Phys.J. {\bf C17}, 623 (2000),
  \href{http://arxiv.org/abs/hep-ph/0007101}{{\tt arXiv:hep-ph/0007101}}.

\bibitem{Feldmann:1998sh}
T.~Feldmann, P.~Kroll and B.~Stech,
\newblock Phys.Lett. {\bf B449}, 339 (1999),
  \href{http://arxiv.org/abs/hep-ph/9812269}{{\tt arXiv:hep-ph/9812269}}.

\bibitem{Feldmann:1998vh}
T.~Feldmann, P.~Kroll and B.~Stech,
\newblock Phys.Rev. {\bf D58}, 114006 (1998),
  \href{http://arxiv.org/abs/hep-ph/9802409}{{\tt arXiv:hep-ph/9802409}}.

\bibitem{Neff:2001zr}
H.~Neff, N.~Eicker, T.~Lippert, J.~W. Negele and K.~Schilling,
\newblock Phys.Rev. {\bf D64}, 114509 (2001),
  \href{http://arxiv.org/abs/hep-lat/0106016}{{\tt arXiv:hep-lat/0106016}}.

\bibitem{Jansen:2009xp}
K.~Jansen and C.~Urbach,
\newblock Comput.Phys.Commun. {\bf 180}, 2717 (2009),
  \href{http://arxiv.org/abs/0905.3331}{{\tt arXiv:0905.3331}}.

\bibitem{R:2005}
{R Development Core Team},
\newblock {\em R: A language and environment for statistical computing},
\newblock R Foundation for Statistical Computing, Vienna, Austria, 2005,
\newblock {ISBN} 3-900051-07-0.

\end{thebibliography}

\end{document}